\documentclass[12 pt,draftcls]{IEEEtran} \onecolumn

\usepackage[dvips]{graphicx}
\usepackage[cmex10]{amsmath}
\usepackage{amssymb}
\usepackage{algorithmic}
\usepackage{array}
\usepackage{mdwmath}
\usepackage{eqparbox}
\usepackage[tight,footnotesize]{subfigure}
\usepackage{graphicx}
\usepackage{algorithmic}
\usepackage{algorithm}
\usepackage{paralist}
\usepackage{float}
\usepackage{color}
\usepackage{wrapfig}
\usepackage{cite}
\usepackage{multirow}

\DeclareMathOperator \Slope{slope}

\title{Sensor Selection for Target Tracking in Wireless Sensor Networks with Uncertainty}

\author{Nianxia Cao, {\it Student Member}, {\it IEEE}, Sora Choi, \\Engin Masazade, {\it Member}, {\it IEEE}, Pramod~K.~Varshney, {\it Fellow}, {\it IEEE}

\thanks{N. Cao, S. Choi, and P. K. Varshney are with the Department of Electrical Engineering and Computer Science, Syracuse University, Syracuse, NY, 13244, USA, email: \{ncao, schoi101, varshney\}@syr.edu}
\thanks{E. Masazade is with Department of Electrical and Electronics Engineering, Yeditepe University, Istanbul,  34755,  Turkey, email: engin.masazade@yeditepe.edu.tr}

\thanks{The work of N. Cao, S. Choi, and P. K. Varshney was supported by U.S. Air Force Office of Scientific
Research (AFOSR) under Grant No. FA9550-10-1-0458. The work
of E. Masazade was supported by the Scientific and Technological Research
Council of Turkey (TUBITAK) under Grant 113E220. A preliminary version of this paper appears in the IEEE International Conference on Information Fusion 2013.}
}

\begin{document}
\maketitle

\begin{abstract}
\boldmath In this paper, we propose a multiobjective optimization framework for the sensor selection problem in uncertain Wireless Sensor Networks (WSNs). The uncertainties of the WSNs result in a set of sensor observations with insufficient information about the target. We propose a novel mutual information upper bound (MIUB) based sensor selection scheme, which has low computational complexity, same as the Fisher information (FI) based sensor selection scheme, and gives estimation performance similar to the mutual information (MI) based sensor selection scheme. 
Without knowing the number of sensors to be selected \textit{a priori}, the multiobjective optimization problem (MOP) gives a set of sensor selection strategies that reveal different trade-offs  between two conflicting objectives: minimization of the number of selected sensors and minimization of the gap between the performance metric (MIUB and FI) when all the sensors transmit measurements and when only the selected sensors transmit their measurements based on the sensor selection strategy. Illustrative numerical results that provide valuable insights are presented. 
\end{abstract}

\begin{IEEEkeywords}
Target tracking, sensor selection, Fisher information, mutual information, information fusion, multiobjective optimization, wireless sensor networks.
\end{IEEEkeywords}

\section{Introduction}

A wireless sensor network (WSN) is composed of a large number of densely deployed sensors, where sensors are battery-powered devices with limited signal processing capabilities. When programmed and networked properly, WSNs are very useful in many application areas including battlefield surveillance \cite{bokareva2006wireless}, environment monitoring and target tracking \cite{yick2008wireless}, industrial processes \cite{gungor2009industrial} and health monitoring and control \cite{milenkovic2006wireless}. 
In our work presented in this paper, the task of the WSN is to track a target emitting or reflecting energy in a given region of interest (ROI), and the sensors send their observations regarding the target to a central node called the fusion center, which is responsible for the final inference.

Target tracking problems often require coverage of broad areas and a large number of sensors that can be densely deployed over the ROI. This results in new challenges when the resources (bandwidth and energy) are limited. In such situations, it is inefficient to utilize all the sensors in the ROI including the uninformative ones, which hardly contribute to the tracking task at hand but still consume resources. This issue has been investigated and addressed via the development of sensor selection schemes, whose goal is to select the best non-redundant set of sensors for the tracking task while satisfying some performance and/or resource constraints \cite{rowaihy2007survey}. 
The sensor selection problem for target localization and target tracking has been considered in  \cite{wang2004entropy,williams2007approximate,
ac10:hoffman,Zhao_spm02,long_analog,
long_quantized,engin_localization,Joshi:tsp09,
Mo&Sinopoli_automatica,Sliu_tsp15,Sliu_tsp14} among others, where the sensor sets are selected to get the desired information gain or reduction in estimation error about the target state.
In \cite{wang2004entropy,williams2007approximate,
ac10:hoffman,Zhao_spm02}, the mutual information (MI) or entropy is considered as the performance metric, and in \cite{long_analog,long_quantized}, the sensors that have the lowest posterior Cramer-Rao lower bound (PCRLB), which is the inverse of the Fisher information (FI), are selected. In \cite{engin_localization}, the authors compared the two sensor selection criteria namely MI and PCRLB for the sensor selection problem based on quantized data, and showed that the PCRLB based sensor selection scheme achieves similar mean square error (MSE) with significantly less computational effort. In \cite{Joshi:tsp09}, the sensor selection problem was formulated as an integer programming problem, which was relaxed and solved through convex optimization. In \cite{Mo&Sinopoli_automatica}, a multi-step sensor selection strategy by reformulating the Kalman filter was proposed, which was able to address different performance metrics and constraints on available resources. In \cite{Sliu_tsp15}, the authors aimed to find the optimal sparse collaboration
topologies subject to a certain information or energy constraint in the context of distributed estimation. For a more complete literature review on sensor management for target tracking, see \cite{Sliu_tsp14} and references therein.

The previous research on sensor selection assumes that the WSNs operate reliably during the target tracking process without any interruptions. 
The fact is that, in some situations, the sensor observations are quite uncertain \cite{nahi1969optimal,hadidi1979linear,
hounkpevi2007robust,
zhang2011robust,xu2006jamming}. For example, sensors may have temporary failure, there may be abrupt changes in the operating environment \cite{hounkpevi2007robust,zhang2011robust}, or other interference such as traffic or birds/animals that may change the power received by the sensors. Moreover, some random interruptions may appear over the communication channels in the system, and adversaries may jam wireless communications using different attack strategies \cite{xu2006jamming}. These types of uncertainties would result in the set of sensor observations with insufficient information about the target at the fusion center.
In other words, in such an uncertain WSN, sensor observations may contain useful information regarding the target only with a certain probability. 
It is important to investigate the sensor selection problem in such an uncertain environment. In our work here, we study the uncertainty caused by occlusions, i.e., the sensors may not be able to observe the target when blocked by some obstacles. 
Regarding the representation of this type of  uncertainty, the authors in \cite{nahi1969optimal} and \cite{hadidi1979linear} introduced a stochastic model for sensor measurements.  Furthermore, the work in \cite{hounkpevi2007robust} and \cite{zhang2011robust} generalized the model in \cite{nahi1969optimal,hadidi1979linear} to multiple sensors by considering a more realistic viewpoint in that the sensors have different uncertainty at different time instants. For the problems involving uncertain WSNs, even though there are studies about the Kalman filter for target tracking
\cite{hadidi1979linear,mariton1990jump,
costa2002stationary,sinopoli2004kalman}, and about the target localization problem with non-ideal channels \cite{ozdemir2009channel,masazade2010channel}, the sensor selection problem in WSNs with uncertain sensor observations has not been considered in the literature and is the subject of this paper.   

In the aforementioned literature, sensor selection schemes require \textit{a priori} information about the number of sensors to be  selected at each time, denoted as $A$, and computationally efficient algorithms are developed in order to find the optimal $A$ sensors that achieve the maximum performance gain. 
Realistically, in many applications like target tracking, it is unlikely that the number of sensors that need to be selected at each time step of tracking is known to the system designer before operation begins. 
Therefore, it is quite necessary and important to investigate sensor selection strategies that determine the optimal number of sensors to be selected as well as which sensors to select based on the WSN conditions. 

Sensor network design usually involves consideration of multiple conflicting objectives, such as maximization of the lifetime of the network or the inference performance, while minimizing the cost of resources such as energy, communication or deployment costs \cite{engin_MOP,ramesh:EC05,ramesh:wamicon10,
nasir:CEC12}. The problems that investigate the trade-offs among such conflicting objective functions are called Multiobjective Optimization Problems (MOPs). In our preliminary work \cite{ncao_Fusion13}, a sensor selection method utilizing FI as the performance metric in an MOP framework has been presented under the assumption that the sensors in the WSN are all reliable. Our work in \cite{ncao_Fusion13} optimized two objectives simultaneously: minimization of the total number of sensors selected at each time, and minimization of the information gap between the FI when all the sensors transmit their measurements and the FI when only the selected sensors transmit their measurements.  
In this work, we investigate the sensor selection problem in an uncertain WSN, and generalize the approach presented in \cite{ncao_Fusion13} by addressing the issues that arise due to uncertainty. As we will see in the paper, the FI based selection scheme (FISS) tends to select sensors which are relatively close to the target, while the MI based selection scheme (MISS) selects sensors that have high sensing probabilities, and achieves better performance. 
The better performance of MISS comes along with high computational complexity. Thus, we propose to use a mutual information upper bound (MIUB) as the performance metric for the sensor selection problem. The complexity of computing MIUB is similar to that of evaluating FI, and is much lower than that of computing MI. We also show through simulation experiments that the MIUB based selection scheme (MIUBSS) hardly degrades the tracking performance. Furthermore, we consider our sensor selection problem with uncertainty under the MOP framework, where the Nondominating Sorting Genetic Algorithm-II (NSGA-II) is applied to dynamically select an optimal set of sensors at each time step. Numerical results show that MIUBSS selects more sensors than FISS under the MOP framework. We also compare our framework with some other sensor selection methods, e.g., weighted sum method and convex optimization method, and show that NSGA-II with the compromise solution (to be discussed later in the paper) 
adaptively decides the optimal number of sensors at each time step of tracking and achieves satisfactory estimation performance while obtaining savings in terms of number of sensors.


The rest of this paper is organized as follows: In Section \ref{Sec:Sys_Model}, we introduce the uncertain WSN system model. The target tracking framework using a particle filter is given in Section \ref{Sec:pf}. In Section \ref{Sec:information}, the performance metric, FI and MI, for sensor selection are introduced, and comparisons are performed through numerical experiments.  Then in Section \ref{Sec:MOP}, we review the fundamentals of MOP and apply NSGA-II to solve the proposed MOP; we also investigate the performance of the MOP framework through simulations in this section. Section \ref{Sec:Concl} is devoted to our conclusions and future research directions.

\section{System model}
\label{Sec:Sys_Model}

We consider a target tracking problem, where a moving target emitting (or reflecting) a signal over an area of interest is tracked by a WSN consisting of $N$ sensors. The target state is assumed to be a 4-dimensional vector $\mathbf{x}_t = [x_t, y_t, \dot{x}_t, \dot{y}_t]$ where $x_t$ and $y_t$ are the target positions, and $\dot{x}_t$ and $\dot{y}_t$ are the target velocities in the horizontal and vertical directions. Even though the approaches developed in this paper are applicable to more complex dynamic models, here 
we assume a linear dynamic model
\begin{equation}\label{eq:model_linear}
\mathbf{x}_{t+1} = \mathbf{F}\mathbf{x}_t + \mathbf{w}_t.
\end{equation}
where $F$ is the state transition matrix and $\mathbf{w}_t$ is the Gaussian process noise with zero mean 
and covariance matrix $Q$: 
\begin{equation}\label{eq:model_FQ}
\mathbf{F}= \left[ \begin{array}{cccc}
1 & 0 & \mathcal{D} & 0 \\
0 & 1 & 0 & \mathcal{D} \\
0 & 0 & 1 & 0 \\
0 & 0 & 0 & 1 \end{array} \right],
\mathbf{Q}= q \left[ \begin{array}{cccc}
\frac{\mathcal{D}^3}{3} & 0 & \frac{\mathcal{D}^2}{2} & 0 \\
0 & \frac{\mathcal{D}^3}{3} & 0 & \frac{\mathcal{D}^2}{2} \\
\frac{\mathcal{D}^2}{2} & 0 & \mathcal{D} & 0 \\
0 & \frac{\mathcal{D}^2}{2} & 0 & \mathcal{D} \end{array} \right],
\end{equation}
where ${\cal D}$ is the sampling interval and $q$ is the process noise parameter. It is assumed that the signal emitted by the target follows  a power attenuation model \cite{ozdemir2009channel}.  
Thus, the signal power received by sensor $i$ which is located at $(x_i,y_i)$ is 
\begin{equation}\label{eq:power_attenuation}
P_{i,t}(\mathbf{x}_t) = \dfrac{P_0}{1+\alpha d_{i,t}^n}
\end{equation}
where $P_0$ is the emitted signal power from the target at distance zero, $n$ is the signal decay exponent and $\alpha$ is a scaling parameter. In \eqref{eq:power_attenuation}, $d_{i,t}$ is the distance between the target and the $i^{th}$ sensor at time step $t$, i.e., $d_{i,t} = \sqrt{(x_t-x_i)^2 + (y_t-y_i)^2}$. 

\subsection{Uncertainty Model of Sensor Observations}
As discussed earlier, sensor observations may be uncertain due to sensor failures, natural interference or some random interruptions. 
Regarding different uncertainties, there are different probabilistic models \cite{lee2010uncertainty}.  
In this paper, we consider the scenario that the sensor observation uncertainty is caused by some obstacles, and assume the following probabilistic measurement model, which has been proposed in \cite{nahi1969optimal} and generalized in \cite{hounkpevi2007robust} and \cite{zhang2011robust}: the sensor observation is assumed to contain only noise if the sensor cannot sense the target due to obstacles, and since such uncertainty may happen at any time for any sensor, the sensing probability may not be identical across the sensors in the WSN, i.e., 
\begin{equation}\label{eq:obs_model}
z_{i,t}=\left\{ \begin{array}{ll}
  h_{i,t}(\mathbf{x}_t) + v_{i,t}, &\mbox{with probability $p_s^{(i)}$} \\
  v_{i,t}, &\mbox{with probability $1-p_s^{(i)}$}\\
       \end{array} \right.
\end{equation}
where $p_s^{(i)}$ is the \textit{sensing probability} of sensor $i$, $h_{i,t}(\mathbf{x}_t) = \sqrt{P_{i,t}(\mathbf{x}_t)}$ represents the signal amplitude received by sensor $i$ at time step $t$, and $v_{i,t}$ is the measurement noise, which is assumed to be independent across time steps and across sensors, follows a Gaussian distribution with parameters ${\cal N}(0,\sigma^2)$. The likelihood function for sensor measurements $\mathbf{z}_t = [z_{i,t},\ldots,z_{N,t}]^T$ given the target state $\mathbf{x}_t$ is simply the product of each sensor $i$'s likelihood function.
Given $\mathbf{x}_t$, $z_{i,t}$ follows the  Gaussian distribution ${\cal N}(h_{i,t}(\mathbf{x}_t), \sigma^2)$ with probability $p_s^{(i)}$, and follows the Gaussian distribution ${\cal N}(0, \sigma^2)$ with probability $1-p_s^{(i)}$, i.e., 
\begin{equation}\label{eq:AnalogModel}
p(z_{i,t}|\mathbf{x}_t) = p_s^{(i)} ~{\cal N}(h_{i,t}(\mathbf{x}_t), \sigma^2) + (1-p_s^{(i)})~{\cal N}(0, \sigma^2)
\end{equation}

For communication between the fusion center and the sensors, we consider the following two practical scenarios: \begin{inparaenum}
\item the sensors directly send their analog measurements $\mathbf{z}_t$ to the fusion center;
and,
\item the sensors quantize their analog measurements to $M$ bits, and then transmit the quantized data to the fusion center for tracking.
\end{inparaenum} Analog sensor measurements contain complete information about the observation, at the expense of high communication cost; on the other hand, quantized measurements save communication burden, but lose some information about the target.

The quantized measurement of sensor $i$ at time step $t$, $D_{i,t}$, is defined as:
\begin{equation}
D_{i,t} = \left\{ \begin{array}{rl}
  0 &\mbox{ $\eta_0 \leq z_{i,t} \leq\eta_{1}$} \\
  1 &\mbox{ $\eta_{1} \leq z_{i,t} \leq \eta_{2}$}\\
\vdots\\
  L-1 &\mbox{ $\eta_{(L-1)} \leq z_{i,t} \leq\eta_L$ }  \\
       \end{array} \right.
\end{equation}where $\boldsymbol{\eta} = [\eta_{0}, \eta_{1},\ldots,\eta_{L}]^T$ is the set of quantization thresholds with $\eta_{0} = -\infty$ and $\eta_{L} = \infty$ and $L=2^M$ is the number of quantization levels. 
The probability that $D_{i,t}$ takes the value $l$ is 
\begin{equation}\label{eq:QuantizeModel}
\begin{aligned}
&p(D_{i,t}=l|\mathbf{x}_t) = \Pr(\eta_l\leq z_{i,t} \leq \eta_{l+1}|\mathbf{x}_t)\\
&= p_s^{(i)}~\Pr(\eta_l\leq z_{i,t} \leq \eta_{l+1}| z_{i,t} \sim {\cal N}(h_{i,t}(\mathbf{x}_t), \sigma^2))\\
&\quad + (1-p_s^{(i)})~\Pr(\eta_l\leq z_{i,t} \leq \eta_{l+1}| z_{i,t} \sim {\cal N}(0, \sigma^2))\\
&= p_s^{(i)} ~ \Big[Q\left(\frac{\eta_{l}-h_{i,t}(\mathbf{x}_t)}{\sigma}\right)-Q\left(\frac{\eta_{l+1}-h_{i,t}(\mathbf{x}_t)}{\sigma}\right)\Big]\\
&\quad + (1-p_s^{(i)})~\Big[Q\left(\frac{\eta_{l}}{\sigma}\right)-Q\left(\frac{\eta_{l+1}}{\sigma}\right)\Big]
\end{aligned}
\end{equation} 
where $Q(\cdot)$ denotes the complementary distribution of the standard Gaussian distribution with zero mean and unit variance
\begin{equation}
Q(x) = \int_{x}^{\infty} \frac{1}{\sqrt{2\pi}} \exp\{-\frac{y^2}{2}\} \mathrm{d} y
\end{equation} Since the sensor measurements are conditionally independent, the likelihood function of $\mathbf{D}_t = [D_{1,t}, D_{2,t},...,D_{N,t}]^T$ can be written as the product of each sensor $i$'s likelihood function.


\section{Particle Filtering for Target Tracking}
\label{Sec:pf}
\begin{algorithm} [tb]                     
\caption{SIR Particle Filter for target tracking}          
\label{alg1}                           
\begin{algorithmic}[1]                    
\STATE Set $t = 1$. Generate initial particles $\mathbf{x}_{0}^{s} \sim p(\mathbf{x}_0)$ with $\forall s\;, w_0^s = N_s^{-1}$.
\WHILE {$t \leq T_s$}
\STATE $\mathbf{x}_{t}^{s} = \mathbf{F}\mathbf{x}_{t-1}^{s} +  \boldsymbol{\upsilon}_t$ (Propagating particles)
\STATE $p(\mathbf{x}_{t}|\mathbf{z}_{1:t}) = \frac{1}{N_s}\sum_{s=1}^{N_s} \delta(\mathbf{x}_{t} - \mathbf{x}_{t}^{s})$
\STATE Obtain sensor data $\mathbf{z}_{t}$ 
\STATE $w_{t}^{s} \propto  p(\mathbf{z}_{t}|\mathbf{x}_{t}^{s})$ (Updating weights through obtained data)
\STATE $w_{t}^{s} = \frac{w_{t}^{s}}{\sum_{s=1}^{N_s} w_{t}^{s}}$ (Normalizing weights)
\STATE $\mathbf{\hat{x}}_{t} = \sum_{s=1}^{N_s} w_{t}^{s} \mathbf{x}_{t}^{s}$
\STATE $\{\mathbf{x}_{t}^{s},N_s^{-1}\} = \textrm{Resampling}(\mathbf{x}_{t}^{s},w_{t}^{s}) $
\STATE $t = t+1$
  \ENDWHILE
\end{algorithmic}
\end{algorithm}

The target tracking problem requires the estimation of the target state using a sequence of sensor measurements. For nonlinear systems, extended Kalman filter (EKF) provides suboptimal solutions. However, 
when the sensor measurements are quantized, even for linear and Gaussian systems, the EKF fails to provide an acceptable performance especially when the number of quantization levels is small \cite{Willet:aes08}. Thus, we employ a sequential importance resampling (SIR) particle filter to solve our nonlinear target tracking problem with analog and quantized sensor measurements \cite{gordon1993novel,particle}. The SIR algorithm is based on the Monte Carlo method, and can be used for recursive Bayesian filtering problems under very weak assumptions \cite{particle}. 
The main idea of the particle filter is to find a discrete representation of the posterior distribution $p(\mathbf{x}_{t}|\mathbf{z}_{1:t})$ ($p(\mathbf{x}_{t}|\mathbf{D}_{1:t})$) by using a set of particles $\mathbf{x}_t^s$ with associated weights $w_t^s$,
\begin{equation}
\label{eq:particle_filter_original}
p(\mathbf{x}_{t}|\mathbf{z}_{1:t}) \approx \sum_{s=1}^{N_s} w_t^s \delta(\mathbf{x}_{t} - \mathbf{x}_{t}^s),
\end{equation} where, $\delta(\cdot)$ is the Dirac delta measure, and $N_s$ denotes the total number of particles. When the number of particles is large enough, the weighted sum of the particles based on the Monte Carlo characterization will be an equivalent representation of the posterior distribution. The resampling step in the SIR particle filter avoids the situation that all but one of the importance weights are close to zero after a few iterations, which is known as the degeneracy phenomenon in the particle filter. Algorithm \ref{alg1} provides a summary of the SIR particle filtering algorithm for the target tracking problem with analog data $\mathbf{z}_t$, where $T_s$ denotes the number of time steps over which the target is tracked, and $\mathbf{z}_t$ is replaced by $\mathbf{D}_t$ if quantized data is utilized for transmission. 

\section{Sensor Selection Criteria for Uncertain WSNs}
\label{Sec:information}
In this section, we present and investigate three performance metrics, FI, MI, and MIUB, for the sensor selection problem in an uncertain WSN. After formulating the three performance metrics mathematically for the analog data and quantized data respectively, we compare them with respect to the resulting tracking performance. 

\subsection{Fisher Information}
Posterior Cramer-Rao Lower Bound (PCRLB) provides the theoretical performance limit for a Bayesian estimator \cite{trees2001modulation}. Let $p(\mathbf{z}_t,\mathbf{x}_t)$ denote the joint probability density function of the sensor measurements and the target state, and let $\hat{\mathbf{x}}_t$ denote the estimate of $\mathbf{x}_t$. The PCRLB on the estimation error is represented as \cite{trees2001modulation},
\begin{equation}\label{pcrlb}
E\left\{[\hat{\mathbf{x}}_t - \mathbf{x}_t][\hat{\mathbf{x}}_t - \mathbf{x}_t]^T\right\} \geq J_t^{-1}
\end{equation}
where $J_t$ is the Fisher information (FI) matrix. It has been shown in \cite{engin_bit_allocation} that, the FI matrix for Bayesian estimation is composed of two parts: the FI obtained from the sensor measurements and the FI corresponding to \textit{a priori} information. Furthermore, under the assumption that the sensor measurements are conditionally independent given the target state $\mathbf{x}_t$, the FI obtained from the measurements of multiple sensors can be written as the summation of each sensor's FI plus the FI from the prior information,
\begin{equation}
J_t \triangleq \sum_{i=1}^N\int_{\mathbf{x}_t}  J_{i,t}^S (\mathbf{x}_t) p(\mathbf{x}_t) d \mathbf{x}_t + J_t^P 
\end{equation}
where $J_t^P$ is the FI matrix of the \textit{a priori} information, and $J_{i,t}^S(\mathbf{x}_t)$ represents the standard FI of each sensor as a function of the target state $\mathbf{x}_t$,
\begin{equation}\label{eq:JS}
\begin{aligned}
J_{i,t}^S(\mathbf{x}_t) &= \int_{z_{i,t}} \dfrac{1}{p(z_{i,t}|\mathbf{x}_t)} \Big(\dfrac{\partial p(z_{i,t}|\mathbf{x}_t)}{\partial \mathbf{x}_t}\Big)^2 \mathrm{d} z_{i,t}
\end{aligned}
\end{equation}

\subsubsection{Fisher information for the analog sensor measurement model}
The FI for analog data is obtained by substituting the likelihood function $p(z_{i,t}|\mathbf{x}_t)$ given in \eqref{eq:AnalogModel} into \eqref{eq:JS}. The derivative of $p(z_{i,t}|\mathbf{x}_t)$ is
\begin{equation}\label{eq:partial_pz}
\begin{aligned}
&\quad\dfrac{\partial p(z_{i,t}|\mathbf{x}_t)}{\partial \mathbf{x}_t} \\
&= p_s^{(i)} ~\dfrac{z_{i,t}-h_{i,t}(\mathbf{x}_t)}{\sigma^2\sqrt{2\pi \sigma^2} } ~\exp \Big\{-\dfrac{(z_{i,t}-h(\mathbf{x}_t))^2}{2\sigma^2}\Big\} \dfrac{\partial h_{i,t}(\mathbf{x}_t)}{\partial \mathbf{x}_t}
\end{aligned}
\end{equation}
where
\begin{equation}\label{eq:partial_model}
\dfrac{\partial h_{i,t}(\mathbf{x}_t)}{\partial \mathbf{x}_t} = \frac{\alpha n}{2} \dfrac{h_{i,t}(\mathbf{x}_t) d_{i,t}^{n-2}}{1+\alpha d_{i,t}^n} \left[\begin{array}{c}
x_t-x_i\\y_t-y_i\\0\\0
\end{array}\right]
\end{equation}
Substituting \eqref{eq:AnalogModel}, \eqref{eq:partial_pz}, and \eqref{eq:partial_model} into \eqref{eq:JS} and letting $J_{i,t}^{SA}(\mathbf{x}_t)$ denote the standard FI matrix for analog data, $J_{i,t}^{SA}(\mathbf{x}_t)$ is obtained as follows:
\begin{equation}\label{eq:JSA}
\begin{aligned}
&J_{i,t}^{SA}(\mathbf{x}_t) = (p_s^{(i)})^2~ \kappa_{i,t}^A(\mathbf{x}_t) \Big(\dfrac{\partial h_{i,t}(\mathbf{x}_t)}{\partial \mathbf{x}_t}\Big) \Big(\dfrac{\partial h_{i,t}(\mathbf{x}_t)}{\partial \mathbf{x}_t}\Big)^T\\
&= \kappa_{i,t}^A(\mathbf{x}_t) \dfrac{(p_s^{(i)})^2 \alpha^2 n^2 h^2_{i,t}(\mathbf{x}_t) d_{i,t}^{2n-4}}{4(1+\alpha d_{i,t}^n)^2}\\
&\quad \times  \left[\begin{array}{cccc}
	(x_i-x_t)^2  & (x_i-x_t)(y_i-y_t) & 0 & 0\\
	(x_i-x_t)(y_i-y_t)   & (y_i-y_t)^2 & 0 & 0\\
	0 & 0 & 0 & 0\\
	0 & 0 & 0 & 0\\
\end{array}\right] 
\end{aligned}
\end{equation}
where
\begin{equation}
\begin{aligned}
&\kappa_{i,t}^A(\mathbf{x}_t)= \\
&\int \limits_{z_{i,t}}\hspace{-0.1cm}\dfrac{1}{p(z_{i,t}|\mathbf{x}_t)} \Big[\dfrac{z_{i,t}-h_{i,t}(\mathbf{x}_t)}{\sigma^2\sqrt{2\pi \sigma^2} } \exp \big\{-\dfrac{(z_{i,t}-h(\mathbf{x}_t))^2}{2\sigma^2}\big\}\Big]^2 \mathrm{d} z_{i,t}
\end{aligned}
\end{equation}

\subsubsection{Fisher information for the quantized sensor measurement model}
The FI of quantized data is calculated by replacing the likelihood function $p(z_{i,t}|\mathbf{x}_t)$ given in \eqref{eq:JS} with $p(D_{i,t}|\mathbf{x}_t)$ in \eqref{eq:QuantizeModel}. 
Since the derivative of the likelihood function of the quantized observations is
\begin{equation}\label{eq:partial_quantizemodel}
\begin{aligned}
\dfrac{\partial p(D_{i,t}|\mathbf{x}_t)}{\partial \mathbf{x}_t} &= \dfrac{p_s^{(i)}}{\sigma \sqrt{2 \pi}} \Bigg[ \exp\big\{-\dfrac{(\eta_l-h_{i,t}(\mathbf{x}_t))^2}{2\sigma^2}\big\} \\
&\quad - \exp \big\{-\dfrac{(\eta_{l+1}-h_{i,t}(\mathbf{x}_t))^2}{2\sigma^2}\big\} \Bigg]\dfrac{\partial h_{i,t}(\mathbf{x}_t)}{\partial \mathbf{x}_t}
\end{aligned}
\end{equation}
we derive the FI for quantized data by substituting \eqref{eq:partial_model} into \eqref{eq:partial_quantizemodel} as follows:
\begin{equation}\label{eq:JSQ}
\begin{aligned}
&J_{i,t}^{SQ}(\mathbf{x}_t) = \sum_{D_{i,t}} \dfrac{1}{p(D_{i,t}|\mathbf{x}_t)} \Big(\dfrac{\partial p(D_{i,t}|\mathbf{x}_t)}{\partial \mathbf{x}_t}\Big)^2 \\
&= (p_s^{(i)})^2~ \kappa_{i,t}^Q(\mathbf{x}_t) \Big(\dfrac{\partial h_{i,t}(\mathbf{x}_t)}{\partial \mathbf{x}_t}\Big) \Big(\dfrac{\partial h_{i,t}(\mathbf{x}_t)}{\partial \mathbf{x}_t}\Big)^T\\
&= \kappa^Q_{i,t}(\mathbf{x}_t) \dfrac{(p_s^{(i)})^2 \alpha^2 n^2 h^2_{i,t}(\mathbf{x}_t) d_{i,t}^{2n-4}}{4(1+\alpha d_{i,t}^n)^2} \\
&\quad \times \left[\begin{array}{cccc}
	(x_i-x_t)^2  & (x_i-x_t)(y_i-y_t) & 0 & 0\\
	(x_i-x_t)(y_i-y_t)   & (y_i-y_t)^2 & 0 & 0\\
	0 & 0 & 0 & 0\\
	0 & 0 & 0 & 0\\
\end{array}\right] 
\end{aligned}
\end{equation}
where
\begin{equation}
\begin{aligned}
\kappa^Q_{i,t}(\mathbf{x}_t) &= \sum_{D_{i,t}} \dfrac{1}{2 \pi \sigma^2 p(D_{i,t}|\mathbf{x}_t)} \Big[ \exp\big\{-\dfrac{(\eta_l-h_{i,t}(\mathbf{x}_t))^2}{2\sigma^2}\big\}\\
& \quad - \exp \big\{-\dfrac{(\eta_{l+1}-h_{i,t}(\mathbf{x}_t))^2}{2\sigma^2}\big\}\Big]^2
\end{aligned}
\end{equation}
Thus, we get the FI for the analog observation model in \eqref{eq:JSA}, and for the quantized observation model in \eqref{eq:JSQ}.

\subsection{Mutual Information}
Information-theoretic sensor management for target tracking seeks to minimize the uncertainty in the estimate of the target state conditioned on the sensor measurements \cite{ryan2008information}. Entropy, which is defined by Shannon \cite{shannon2001mathematical}, represents the uncertainty or randomness in the estimate of the target state $\mathbf{x}_t$. Moreover, because of the relationship between the entropy and the MI \cite{cover2012elements}, the sensor selection problem for target tracking can be solved by maximizing the MI between the target state and the sensor measurements. 

Given the distribution of the target state and the likelihood function of the sensor measurements, the MI for the analog data can be written as  \cite{ac10:hoffman,engin_localization}
\begin{equation}\label{eq:analog_MI}
\begin{aligned}
&I(\mathbf{x}_t,\mathbf{z}_t) = H(\mathbf{z}_t) - H(\mathbf{z}_t|\mathbf{x}_t)\\
&= - \int_{\mathbf{z}_t} \Bigg\{\int_{\mathbf{x}_t} p(\mathbf{z}_t|\mathbf{x}_t) p(\mathbf{x}_t) \mathrm{d} \mathbf{x}_t\Bigg\}\\
& \qquad \qquad \Bigg\{\log_2 \bigg[\int_{\mathbf{x}_t} p(\mathbf{z}_t|\mathbf{x}_t) p(\mathbf{x}_t) \mathrm{d} \mathbf{x}_t\bigg]\Bigg\}\mathrm{d} \mathbf{z}_t \\
&\quad + \sum_{i=1}^N \int_{\mathbf{x}_t}\Bigg[\int_{z_{i,t}}  p(z_{i,t}|\mathbf{x}_t)  \log_2 p(z_{i,t}|\mathbf{x}_t) \mathrm{d} z_{i,t}\Bigg] p(\mathbf{x}_t)\mathrm{d} \mathbf{x}_t
\end{aligned}
\end{equation}
where $H(\mathbf{z}_t)$ is the entropy of the sensor measurements $\mathbf{z}_t$, and $H(\mathbf{z}_t|\mathbf{x}_t)$ is the conditional entropy of the sensor measurements $\mathbf{z}_t$ given the target state $\mathbf{x}_t$. Similarly, the MI for the quantized sensor measurements can be written as
\begin{equation}\label{eq:quantize_MI}
\begin{aligned}
&I(\mathbf{x}_t,\mathbf{D}_t) = H(\mathbf{D}_t) - H(\mathbf{D}_t|\mathbf{x}_t)\\
&=- \sum_{\mathbf{D}_t} \Bigg\{\int_{\mathbf{x}_t} p(\mathbf{D}_t|\mathbf{x}_t) p(\mathbf{x}_t) \mathrm{d} \mathbf{x}_t\Bigg\}\\
&\qquad \qquad \Bigg\{\log_2 \bigg[\int_{\mathbf{x}_t} p(\mathbf{D}_t|\mathbf{x}_t) p(\mathbf{x}_t) \mathrm{d} \mathbf{x}_t\bigg]\Bigg\}\\
&\quad + \sum_{i=1}^N \int_{\mathbf{x}_t}\Bigg[\sum_{D_{i,t}}  p(D_{i,t}|\mathbf{x}_t)  \log_2 p(D_{i,t}|\mathbf{x}_t) \Bigg] p(\mathbf{x}_t)\mathrm{d} \mathbf{x}_t
\end{aligned}
\end{equation}
where the summation over $\mathbf{D}_t$ is taken over all possible combinations of the quantized measurements of the set of sensors.


\subsection{Mutual Information Upper Bound (MIUB)}
The computational complexity of evaluating the MI for a set of $A$ sensors increases exponentially with the number of sensors $A$, so that 
it becomes impractical to compute the MI in \eqref{eq:analog_MI} and \eqref{eq:quantize_MI} when the number of sensors to be selected is large \cite{ac10:hoffman} \cite{zhang2010efficient}. The chain rule for the MI is described as follows (we only show the MI for analog data, results for quantized data are similar): 
\begin{equation}
I(\mathbf{z}_t;\mathbf{x}_t) = \sum\limits_{i=1}^N I(z_{i,t};\mathbf{x}_t|z_{i-1,t},\cdots,z_{1,t})
\end{equation}
Since $z_{1,t},\cdots,z_{N,t}$ are conditionally independent given the target state $\mathbf{x}_t$, $z_{i,t} \rightarrow \mathbf{x}_t \rightarrow z_{j,t}$ ($i,j \in \{1,\cdots,N\}$) form a Markov chain, and we have the following data processing inequality \cite{cover2012elements}:
\begin{equation}
\begin{aligned}
I(z_{i,t};\mathbf{x}_t|z_{i-1,t},\cdots,z_{1,t}) &\leq I(z_{i,t};\mathbf{x}_t|z_{i-1,t},\cdots,z_{2,t})\\
\cdots & \leq I(z_{i,t};\mathbf{x}_t)
\end{aligned}
\end{equation}
Thus, $\sum_{i=1}^N I(z_{i,t};\mathbf{x}_t)$ is an upper bound on $I(\mathbf{z}_t;\mathbf{x}_t)$. We use this mutual information upper bound (MIUB) as the performance metric for our sensor selection problem. It can be easily shown that the computational complexity of evaluating MIUB for selecting $A$ out $N$ sensors increases linearly with $A$, which is the same with that of computing FI. 

\subsection{Comparison of Performance Metrics for Sensor Selection by Numerical Experiments}
\label{Subsec:criteria_sims}
\begin{figure}[tb]
\centerline{ \begin{tabular}{c}
\includegraphics[width=.7\columnwidth,height=!]{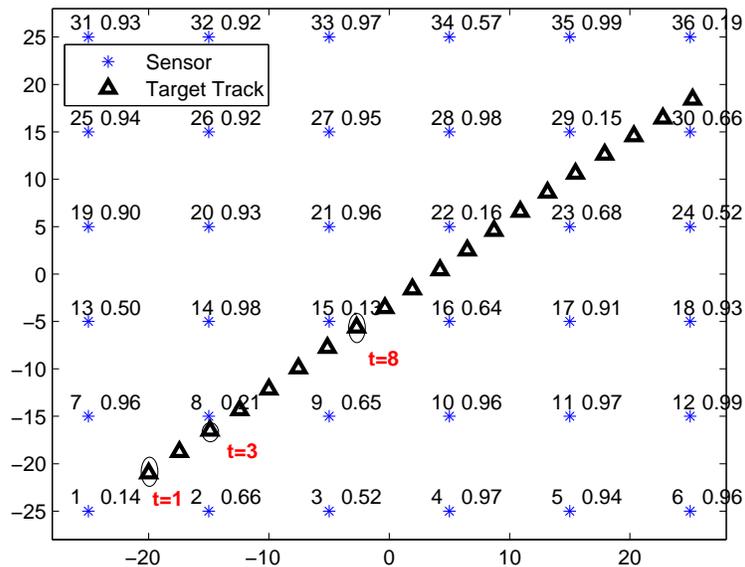} \end{tabular} }
      \caption{WSN with 36 unreliable sensors.  Numbers above the stars indicate sensor index (left) and its sensing probability (right).}
  \label{fig:wsn}
\end{figure}


In this subsection, we compare the performance of the above three performance metrics, FI, MI, and MIUB, for the sensor selection problem through some numerical experiments. 

\paragraph{Simulation setting} In our simulations, we consider the WSN shown in Fig. \ref{fig:wsn}, which has $N=6\times 6 =36$ sensors deployed in the ROI of area $b^2 = 50 \times 50$ $m^2$. In the current work, we assume that the sensing probabilities of the sensors are already known to the fusion center, research on how to learn the probabilities iteratively is an interesting problem and can be considered in the future. Generally, if the sensors around the target tracks have higher sensing probabilities compared to other sensors in the WSN, it is highly likely that the algorithm will select those sensors owing to both higher signal power and sensing probability. Our interest is in considering more challenging cases to test the performance of our algorithm. Thus, we assume that the sensors around the target track have relatively low sensing probabilities as shown in the figure. Moreover, the sensing probabilities may be identical for some sensors if they are in the same environment, however, if the sensors have the same sensing probability, the selection results would be similar to our preliminary work \cite{ncao_Fusion13}. Thus, we consider the scenario in which the sensors in the WSN all have different sensing probabilities. 

For the linear dynamical model of the target given in \eqref{eq:model_linear}, the time interval is ${\cal D} = 1.25$ seconds and the process noise parameter $q=2.5\times 10^{-3}$.  The source power is $P_0 = 1000$ and the variance of the measurement noise is selected as $\sigma = 0.2$. The sensors quantize their observations to $M$ bits for quantized data, and the quantization thresholds $[\eta_1,\cdots,\eta_{L-1}]$ are selected to be the values which evenly partition the interval $[-\sigma,\sigma+\sqrt{P_0}]$. The prior distribution about the state of the target, $p(\mathbf{x}_0)$, is assumed to be Gaussian with mean $\boldsymbol{\mu}_0 = [-23 \hspace{5pt} -24\hspace{5pt} 2\hspace{5pt} 2]^T$ and covariance $\mathbf{\Sigma}_0 = \mathrm{diag}[\sigma^2_{\mathbf{x}}\hspace{5pt} \sigma^2_{\mathbf{x}}\hspace{5pt} 0.01\hspace{5pt} 0.01]$ where we select $\sigma_{\mathbf{x}} = 6$. The initial $N_s = 5000$ particles are drawn from $p(\mathbf{x}_0)$. The mean square error (MSE) is used to measure errors between the ground truth and the estimates, and the MSE of the estimation at each time step of tracking is averaged over $T_{total}$ trials as,
\begin{equation}
\mathrm{MSE}_t = \frac{1}{T_{total}} \sum_{tr=1}^{T_{total}} (\hat{\mathbf{x}}_t^{tr}(1) - \mathbf{x}^{tr}_t(1))^2+(\hat{\mathbf{x}}_t^{tr}(2) - \mathbf{x}^{tr}_t(2))^2
\end{equation}
where $\hat{\mathbf{x}}_t^{tr}$ and $\mathbf{x}_t$ are the estimated and the actual target states at time $t$ of the $tr^{th}$ trial.

\begin{table}
\caption{Sensors with the most significant MI or FI at different time steps}
\label{tab:sensor_index}
\centering
\begin{tabular}{|c|c|c|c|}
\hline
\multirow{2}{*}{Time step} & \multirow{2}{*}{Analog data} & Quantized data & Quantized data\\
& & $M=5$ &  $M=2$\\
\hline\hline
$t=1$ MI & Sensor 2,7 & Sensor 2,7 & Sensor 2,7\\
\hline
$t=1$ FI & Sensor 2,7 & Sensor 2,7 & Sensor 2,7\\
\hline \hline
$t=3$ MI & Sensor 8 & Sensor 8 & Sensor 2,7,14\\
\hline
$t=3$ FI & Sensor 8 & Sensor 8 & Sensor 2,7,9\\
\hline \hline
$t=8$ MI & Sensor 16 & Sensor 16 & Sensor 10,16,21\\
\hline
$t=8$ FI & Sensor 15 & Sensor 15 & Sensor 10,16 \\ \hline
\end{tabular}
\end{table}

\paragraph{Sensors with highest MI or FI at different time steps} We first consider analog and two quantization communication schemes ($M=5$ and 2) for one Monte Carlo run. 
The sensors with highest MI or FI are listed in Table \ref{tab:sensor_index}. 
Note that, 1) since the FI in our paper is a matrix, we consider the determinant of the FI matrix, which corresponds to the area of the uncertainty ellipsoid \cite{bar_shalom_tracking_book}; 2) we are interested in the effect of the sensors' distances from the target and the sensing probabilities on the performance metrics, thus we compute the performance metric for each sensor instead of focusing on different sets of multiple sensors; 3) for individual sensors, the MI and MIUB are identical.

Generally, quantized data contains less information compared with the analog data. We first discuss the results for analog data and 5-bit quantized data. We observe from Table \ref{tab:sensor_index} that the sensors with highest MI or FI are identical for Analog data and 5-bit quantized data, 
which means that 5-bit quantization preserves most information of the analog data as far as sensor selection is concerned. Additionally, we investigate three distinct time steps to compare the results:  
\begin{itemize}
\item At time step 1, the target is relatively close to sensors 2 and 7 with a similar distance from the target, so that sensors 2 and 7 have the most significant MI and FI. Sensors 1 and 8 have very low sensing probabilities though they have similar distance to the target as sensors 2 and 7, and therefore have low MI and FI. 
\item At time step 3, the target is much closer to sensor 8 than the other sensors, so that sensor 8 has the highest MI and FI even though it has a low sensing probability.
\item At time step 8, sensor 15 is the closest one to the target with very low sensing probability, and sensor 16 is the second closest with higher sensing probability. 
In this case, sensor 15 has the highest FI while sensor 16 has the highest MI. 
\end{itemize}  
The 2-bit quantized data contains much less information about the target compared to the analog data and the 5-bit quantized data, so that the sensing probability of sensors affects the FI and MI more with the 2-bit quantized data. Thus, the sensors with relatively higher sensing probabilities have higher FI and MI than the other sensors for the 2-bit quantized data case as shown in Table \ref{tab:sensor_index}.

Therefore, we conclude that for analog data or quantized data with a large number of quantization levels, MI is more affected by the sensing probabilities of the sensors than FI; for quantized data with small number of quantization levels, both MI and FI are considerably affected by the sensing probabilities. Moreover, FISS tends to select sensors which are closer to the target compared to MISS, which can be explained from Equation \eqref{eq:JSA} and \eqref{eq:JSQ} with the corresponding parameters, i.e., the distance between the target and the sensors dominates FI. However, such an explanation cannot be found for MI. In other words, the sensor's distance from the target, sensing probability, and the number of quantization levels are all important factors that determine the tracking performance of the WSNs. 

\begin{figure*}[tb]
\begin{center}
\subfigure[]{
\includegraphics[%
  width=0.45\textwidth, height = !]{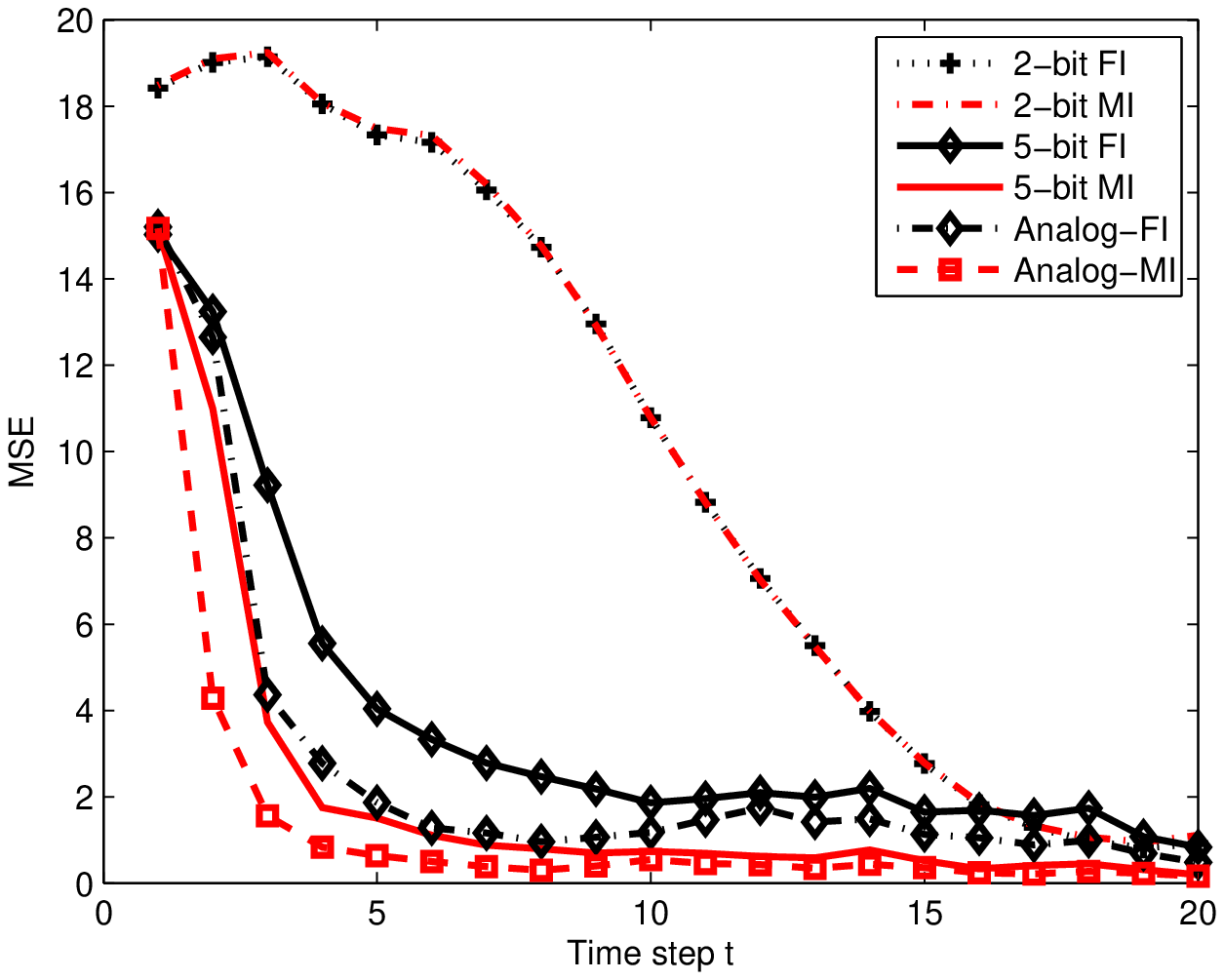}
\label{subfig:mse} }
\subfigure[]{
\includegraphics[%
  width=0.45\textwidth, height = !]{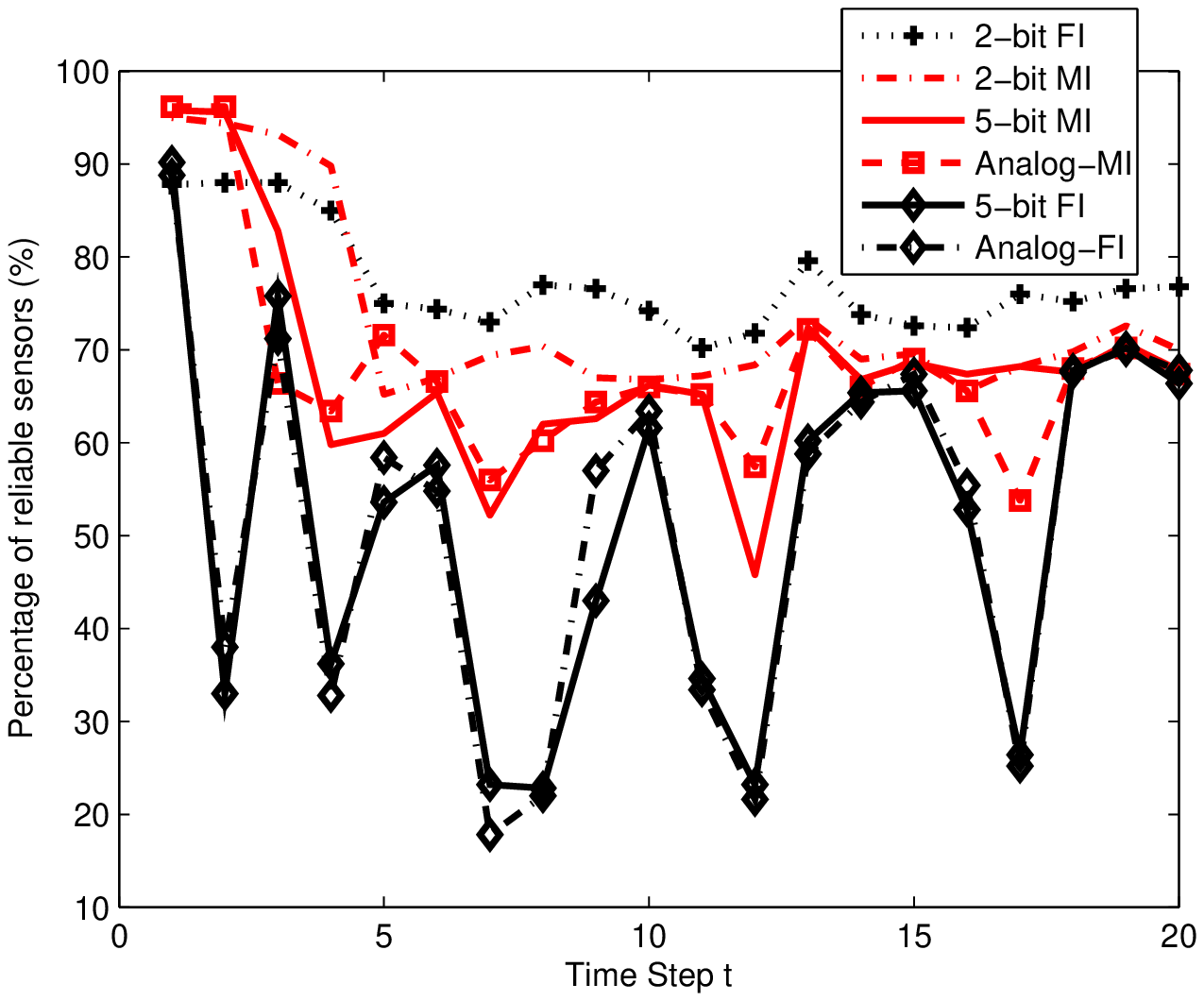}
\label{subfig:reliability} }
\caption{Target tracking performance with analog data, 5-bit quantized data, and 2-bit quantized data, (a) MSE performance; (b) average number of reliable sensors selected.}
\label{fig:onesensor}
\end{center}
\end{figure*}

\paragraph{Tracking performance}
In Fig. \ref{fig:onesensor}, we show the performance of the WSN given in Fig. \ref{fig:wsn} when only one sensor is selected at each time step over $T_{total}=500$ Monte Carlo runs. 
Fig. \ref{subfig:mse} shows that MISS has better MSE performance than FISS with both analog data and 5-bit quantized data. We explain the result by investigating the percentage of \textit{reliable} sensors (the fusion center treats a sensor as \textit{unreliable} if it's amplitude is quite close to noise\footnote{In the experiments, we check if it is within the region $[-3\sigma,3\sigma]$.}) among the selected ones over 500 Monte Carlo trials in Fig. \ref{subfig:reliability}. We observe that, in 500 Monte Carlo trials, around $60\%$ of the sensors selected by MISS are reliable, and only around $40\%$ of the sensors selected by FISS are reliable, which explains the better estimation performance of MISS. Although the sensor selection scheme with 2-bit quantized data selects even more reliable sensors, there is no improvement with respect to the MSE performance because of the significant information loss in the quantization process. As is shown in Fig. \ref{subfig:mse}, the sensor selection scheme based on analog data has the best tracking performance; 5-bit quantized data based sensor selection scheme achieves performance that is close to that with the analog data; and 2-bit quantized data based sensor selection scheme performs much worse. We only show simulation results for the 5-bit quantized data in the following simulation experiments.

\begin{figure}[tb]
\centerline{ \begin{tabular}{c}
\includegraphics[width=.7\columnwidth,height=!]{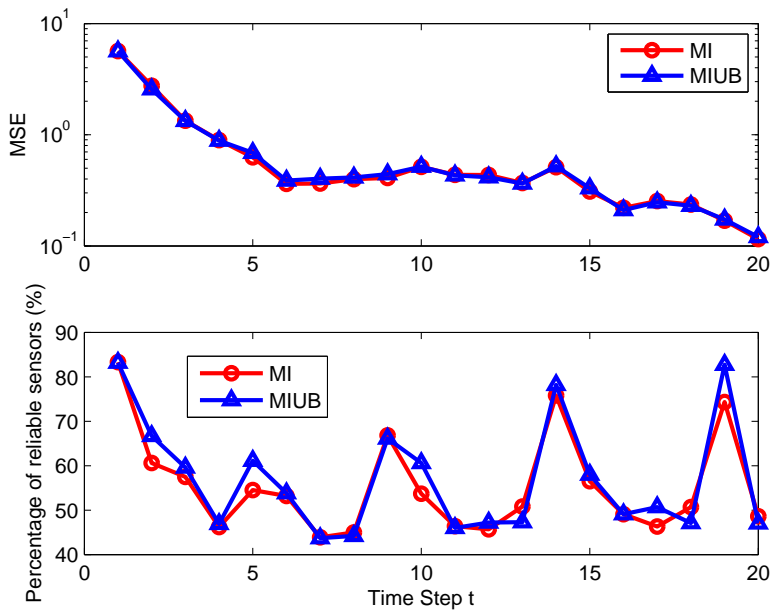} \end{tabular} }
      \caption{Target tracking performance for MI and MIUB, $A=2$.}
  \label{fig:MI_MIUB}
\end{figure}


\paragraph{Performance of MIUBSS} The complexity of computing MIUB for selecting $A$ out of $N$ sensors is the same as that of computing FI (both increase linearly with $A$), and is much less than that of evaluating the MI (increases exponentially with $A$). 
Fig. \ref{fig:MI_MIUB} shows the results of MISS and MIUBSS when $A=2$ sensors are selected, and we observe similar performance for MISS and MIUBSS in terms of both the percentage of reliable sensors selected by the schemes and the MSE performance. 
In other words, MIUBSS obtains performance similar to MISS but with much lower computational complexity. Thus, in the next section, we utilize MIUBSS, instead of MISS, in the multiobjective optimization framework, and compare it with FISS.

\section{Multiobjective Optimization based Sensor Selection}
\label{Sec:MOP}

In this section, we utilize the MOP framework to find the sensor selection strategy that can determine the optimal sensor set. 

The mathematical description of an $n$-objective optimization problem is given as
\begin{equation}
\begin{aligned}
\label{eq:mop_definition}
& \min_{\boldsymbol{\alpha}} & &  \left\{f_1(\boldsymbol{\alpha}),f_2(\boldsymbol{\alpha}),\ldots,f_n(\boldsymbol{\alpha})\right\}\\
& \text{subject to} & & a \leq \alpha_i \leq b, h(\boldsymbol{\alpha}) = 0, g(\boldsymbol{\alpha}) \leq 0
\end{aligned}
\end{equation} where $\boldsymbol{\alpha}$ is the vector of decision variables with elements $\alpha_i$, $a$ and $b$ define the bounds on decision variables, functions $h(.)$ and $g(.)$ represent the equality and inequality constraints of the problem respectively. For the MOP, the solutions satisfying the constraints of (\ref{eq:mop_definition}) form the feasible set $\cal{C}$. In an optimization problem involving the minimization of all the objectives, the solution $\boldsymbol{\alpha}^1$ dominates the solution $\boldsymbol{\alpha}^2$ ($\boldsymbol{\alpha}^1 \succ \boldsymbol{\alpha}^2$) if and only if
\begin{equation}
\begin{aligned}
& f_u(\boldsymbol{\alpha}^1) \leq f_u(\boldsymbol{\alpha}^2) & &  \forall u \in \{ 1,2,\ldots n \} \\
& f_v(\boldsymbol{\alpha}^1) < f_v(\boldsymbol{\alpha}^2) & &  \exists v \in \{ 1,2,\ldots n \}
\end{aligned}
\end{equation}
$\boldsymbol{\alpha}^{\ast}$ is called a Pareto optimal solution if and only if there is no $\boldsymbol{\alpha}$ in $\cal{C}$ that dominates $\boldsymbol{\alpha}^{\ast}$, and the set of Pareto optimal outcomes is called the Pareto front. A well-known technique for solving MOPs is to minimize a weighted sum of the objectives, which yields a single solution corresponding to the weights used. With this approach, if a uniform spread of weights is employed to obtain different solutions, it rarely produces a uniform spread of points on the Pareto front. Some of the optimal solutions may become closely spaced and hence reducing the number of design alternatives \cite{survey_mop}.


In our work, the sensor selection strategies reflect different trade-offs between two objective functions: the estimation performance and the number of selected sensors, which are dependent on the binary decision variables.

\subsection{Objective Functions based on Fisher Information (FI) and Mutual Information Upper Bound (MIUB)}

\subsubsection{FI based objective functions}
Let $\boldsymbol{\alpha}_t = [\alpha_{1,t},\ldots,\alpha_{N,t}]$ be the sensor selection strategy at time step $t$. The elements of $\boldsymbol{\alpha}_t$ are binary variables, i.e, $\alpha_{i,t}=1$, if sensor $i$ is selected and $\alpha_{i,t}=0$ otherwise. Then, $A = \sum_{i=1}^N \alpha_{i,t}$ is the number of sensors selected at time step $t$. Based on the sensor selection strategy $\boldsymbol{\alpha}_t$, the FI matrix at time step $t$ can be written as,
\begin{equation}
J_t (\boldsymbol{\alpha}_t) = \sum_{i=1}^N \alpha_{i,t}J_{i,t}^D + J_t^P
\end{equation} We determine the sensor selection strategy from the solution of the MOP where the objective functions are: minimization of the information gap between the FI based on the measurements of all the sensors and the FI based on the sensor set selected by strategy $\boldsymbol{\alpha}_t$,
\begin{eqnarray}
\label{eq:obj1_fim}
&&f_1(\boldsymbol{\alpha}_t) =  \\
&&\frac{\log{\det\left(\sum_{i=1}^N J_{i,t}^D + J_t^P\right)} - \log{\det\left(\sum_{i=1}^N \alpha_{i,t}J_{i,t}^D + J_t^P\right)}}{\log{\det\left(\sum_{i=1}^N J_{i,t}^D + J_t^P\right)}} \nonumber
\end{eqnarray} and minimization of the normalized number of selected sensors,
\begin{equation}
\label{eq:obj2}
f_2 (\boldsymbol{\alpha}_t) = \frac{1}{N}\sum_{i=1}^N \alpha_{i,t}
\end{equation}

\subsubsection{MIUB based objective functions}
The objective functions based on MIUB are very similar to that with the FI: minimization of the normalized information gap between the total MIUB based on all the sensors and the MIUB based on the sensor selection strategy $\boldsymbol{\alpha}_t$, 
\begin{equation}
f_1(\boldsymbol{\alpha}_t) = \dfrac{\sum_{i=1}^N I^{(i)} - \sum_{i=1}^N \alpha_i I^{(i)}}{\sum_{i=1}^N I^{(i)}}
\end{equation}
where $I^{(i)}$ denotes $I(z_{i,t};\mathbf{x}_t)$ (or $I(D_{i,t};\mathbf{x}_t)$), and minimization of the normalized number of selected sensors (the same as \eqref{eq:obj2}).


\subsection{NSGA-II}
In this paper, we solve the above MOP which has binary decision variables using a state-of-the-art multiobjective evolutionary algorithm, Nondominating sorting genetic algorithm (NSGA)-II \cite{NSGA-II}. This algorithm yields all the solutions on the Pareto front that explore all the possible tradeoffs between conflicting objectives. 

NSGA-II \cite{NSGA-II} first generates an initial population of size $P$ where each solution in the population is a feasible solution of the MOP. In our problem, a solution in the population is represented as a vector of $N$ elements where each element is a binary variable. NSGA-II is an elitist algorithm where good solutions are always preserved in the population. The values of the objective functions for each solution in the population form the fitness values of the solution. Then all the solutions in the population are sorted based on their non-domination. As an example, solutions with Rank 1 consist of all non-dominated solutions, then solutions with Rank 2 consist of all the solutions which are dominated by only one of the solutions in the population and so on. If two solutions in the population have the same fitness value, then they are sorted based on their crowding distance, which is a closure measure of each solution to its neighbors.

NSGA-II uses the rank of a solution to create the mating population.  The offspring solutions are generated by using binary tournament selection \cite{NSGA-II}. If both of the selected solutions have the same fitness value, then the solution with larger crowding distance is selected. In our problem where we have binary decision variables, we use a real-parameter recombination operator called uniform crossover (UX), where offspring solutions $c_1$ and $c_2$ are obtained from parent solutions $p_1$ and $p_2$ according to,
\begin{equation}
\begin{aligned}
c_1 = \xi p_1 + (1-\xi) p_2 \hspace{15pt} \\
c_2 = (1-\xi) p_1 + \xi p_2 \hspace{15pt} \\
\end{aligned}
\end{equation}where $\xi$ is defined by a random number $q$ between $[0,1]$\cite{knapsack}
\begin{equation}
\label{eq:rand_number}
\begin{aligned}
\xi = 1 \hspace{15pt} q \leq 0.5 \\
\xi = 0 \hspace{15pt} q > 0.5 \\
\end{aligned}
\end{equation}
Along with UX, the uniform mutation procedure is employed. In uniform mutation, an offspring solution $c_l$ is obtained from the parent solution $p_l$ according to
\begin{equation}
c_l = \delta (1-c_l) + (1-\delta) c_l
\end{equation}where $\delta$ is also determined according to (\ref{eq:rand_number}). Then the new population with all the parents and offsprings are sorted again based on their non-dominance and the population size is decreased to the original population size $P$ by eliminating all the lower rank solutions. Remaining solutions are then fed to a binary tournament selection operator and so on. After several generations $G$, the population will preserve solutions near or on the Pareto optimal front.

\subsection{Solution Selection from the Pareto-optimal Front}

Since NSGA-II provides $P$ non-dominated solutions, it is necessary to select one particular solution from the Pareto-front which can yield the desired trade-off between the conflicting objectives. In \cite{boyd}, the knee of the trade-off curve is introduced as the solution where a small decrease in one objective is associated with a large increase in the other. Let $\boldsymbol{\alpha}^a$ and $\boldsymbol{\alpha}^b$ be two adjacent (neighboring) solutions on the Pareto-optimal front where $f_1(\boldsymbol{\alpha}^a) > f_1(\boldsymbol{\alpha}^b)$ and $f_2(\boldsymbol{\alpha}^a) < f_2(\boldsymbol{\alpha}^b)$. Then we can compute the slope of the curve between solutions $\boldsymbol{\alpha}^a$ and $\boldsymbol{\alpha}^b$ from,
\begin{equation}
\label{eq:knee_pt}
\Slope\{\boldsymbol{\alpha}^b\} = 180-\left[\arctan\left(\frac{f_1(\boldsymbol{\alpha}^a) - f_1(\boldsymbol{\alpha}^b)}{f_2(\boldsymbol{\alpha}^a) - f_2(\boldsymbol{\alpha}^b)}\right)\frac{180}{\pi}\right]
\end{equation} For our problem, we define $\boldsymbol{\alpha}^1$ as the all zero solution where none of the sensors are selected, so that $f_1(\boldsymbol{\alpha}^1) = 1$ and $f_2(\boldsymbol{\alpha}^1) = 0$. Similarly, we define $\boldsymbol{\alpha}^P$ is the all one solution which yields $f_1(\boldsymbol{\alpha}^P) = 0$ and $f_2(\boldsymbol{\alpha}^P) = 1$. We call the Pareto-optimal solution which maximizes (\ref{eq:knee_pt}) as the knee point solution given by,
\begin{equation}
\label{eq:kp_sol}
\boldsymbol{\alpha}_t = \displaystyle \arg \max_{\boldsymbol{\alpha}^{2},\ldots,\boldsymbol{\alpha}^{P}} \Slope\{\boldsymbol{\alpha}^{\rho}\}
\end{equation}where $\boldsymbol{\alpha}^{\rho}$ ($\rho \in \{2,3,\ldots P\}$) represents the solutions on or near the Pareto-optimal front.

\begin{figure*}[tb]
\begin{center}
\subfigure[]{
\includegraphics[%
  width=0.45\textwidth, height = !]{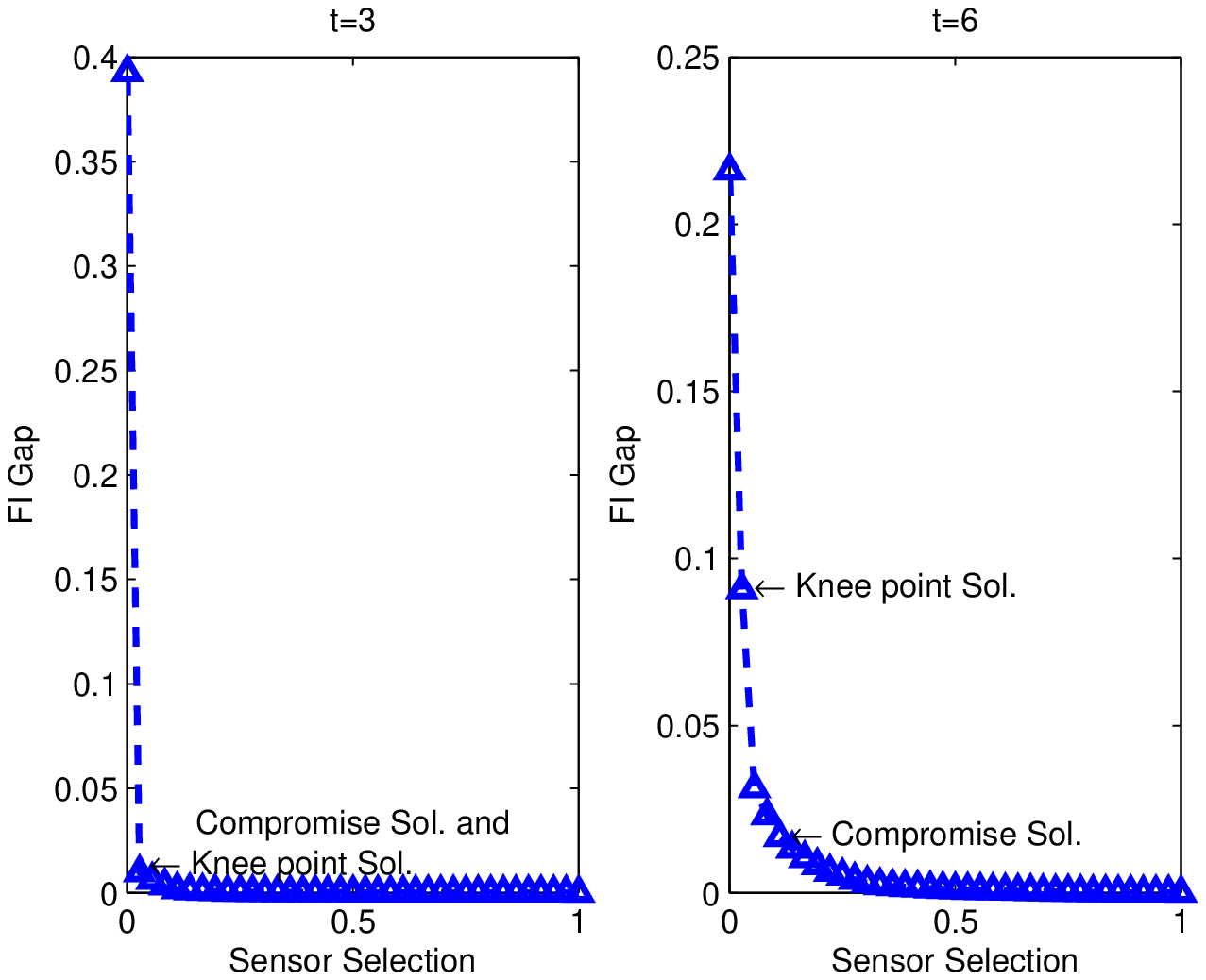}
\label{subfig:front_FI}}
\subfigure[]{
\includegraphics[%
  width=0.45\textwidth, height = !]{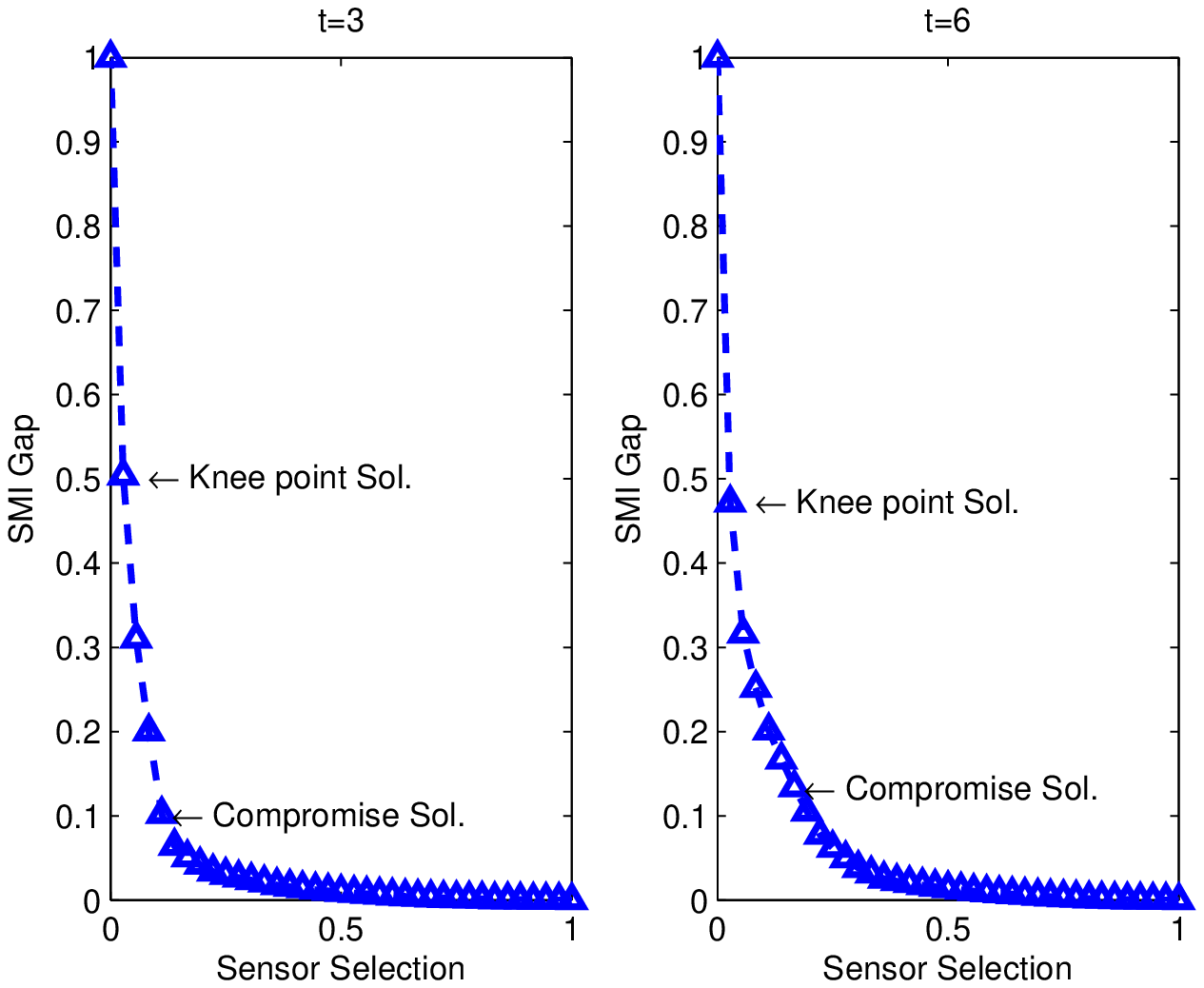}
\label{subfig:front_MIUB}}
\caption{Pareto optimal front obtained by using NSGA-II at time step $t=3$ and $t=6$, (a) FI; (b) MIUB.}
\label{fig:front}
\end{center}
\end{figure*}

Alternatively, the utopia point $F^{\ast}$ of a MOP is defined as \cite{survey_mop},
\begin{equation}
F^{\ast} = \left[ f_1^{\ast},\ldots,f_n^{\ast} \right]^T
\end{equation}where $f_j^{\ast}$ is the individual minima of objective $f_j  \; (j \in \{1,\ldots,n\})$ defined as,
\begin{equation}
f_j^{\ast} = \min_{\boldsymbol{\alpha}} \{ f_j(\boldsymbol{\alpha}) \mid \boldsymbol{\alpha} \in \cal{C}\}
\end{equation} and let $F(\boldsymbol{\alpha}^{\rho}) = \left[ f_1(\boldsymbol{\alpha}^{\rho}),\ldots,f_n(\boldsymbol{\alpha}^{\rho}) \right]^T$ where $\rho \in \{1,2,\ldots,P\}$. In \cite{survey_mop}, the point which is closest to the utopia point has been defined as the compromise solution. In this paper, we use the Euclidean distance to find the compromise solution as,
\begin{equation}
\label{eq:comp_sol}
\boldsymbol{\alpha}_t = \displaystyle \arg \min_{\boldsymbol{\alpha}^{1},\ldots,\boldsymbol{\alpha}^{P}} \sqrt{\sum_{j=1}^n \left(f_{j}^{\ast} - f_{j}(\boldsymbol{\alpha}^{\rho})\right)^2}
\end{equation}In the next section, we present some numerical results.

\subsection{Numerical Experiments for the MOP Framework}
\label{Sec:MOP_sims}

\begin{figure}[t]
\begin{center}
\includegraphics[%
  width=0.7\columnwidth, height = !]{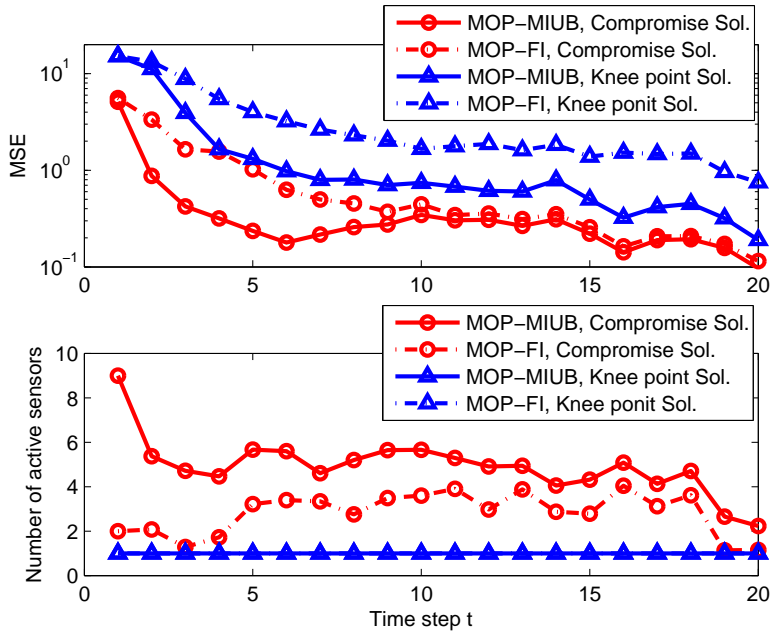}
\caption{Tracking performance at each time step with different solution selection methods.}
\label{fig:MOP_SolS}
\end{center}
\end{figure}

\begin{figure*}[t]
\begin{center}
\subfigure[]{
\includegraphics[%
  width=0.45\textwidth, height = !]{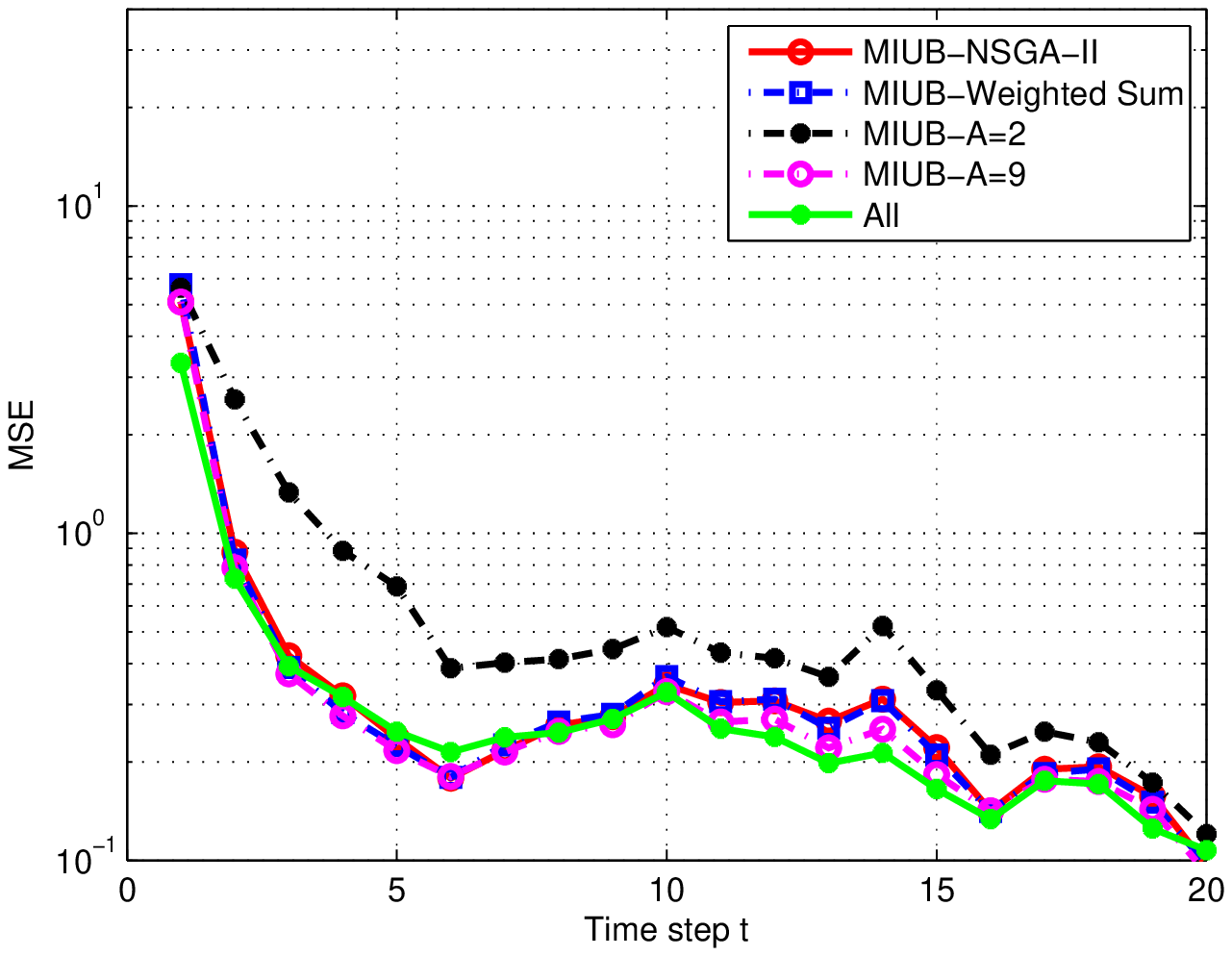}
\label{subfig:MOP_WS_MIUB}}
\subfigure[]{
\includegraphics[%
  width=0.45\textwidth, height = !]{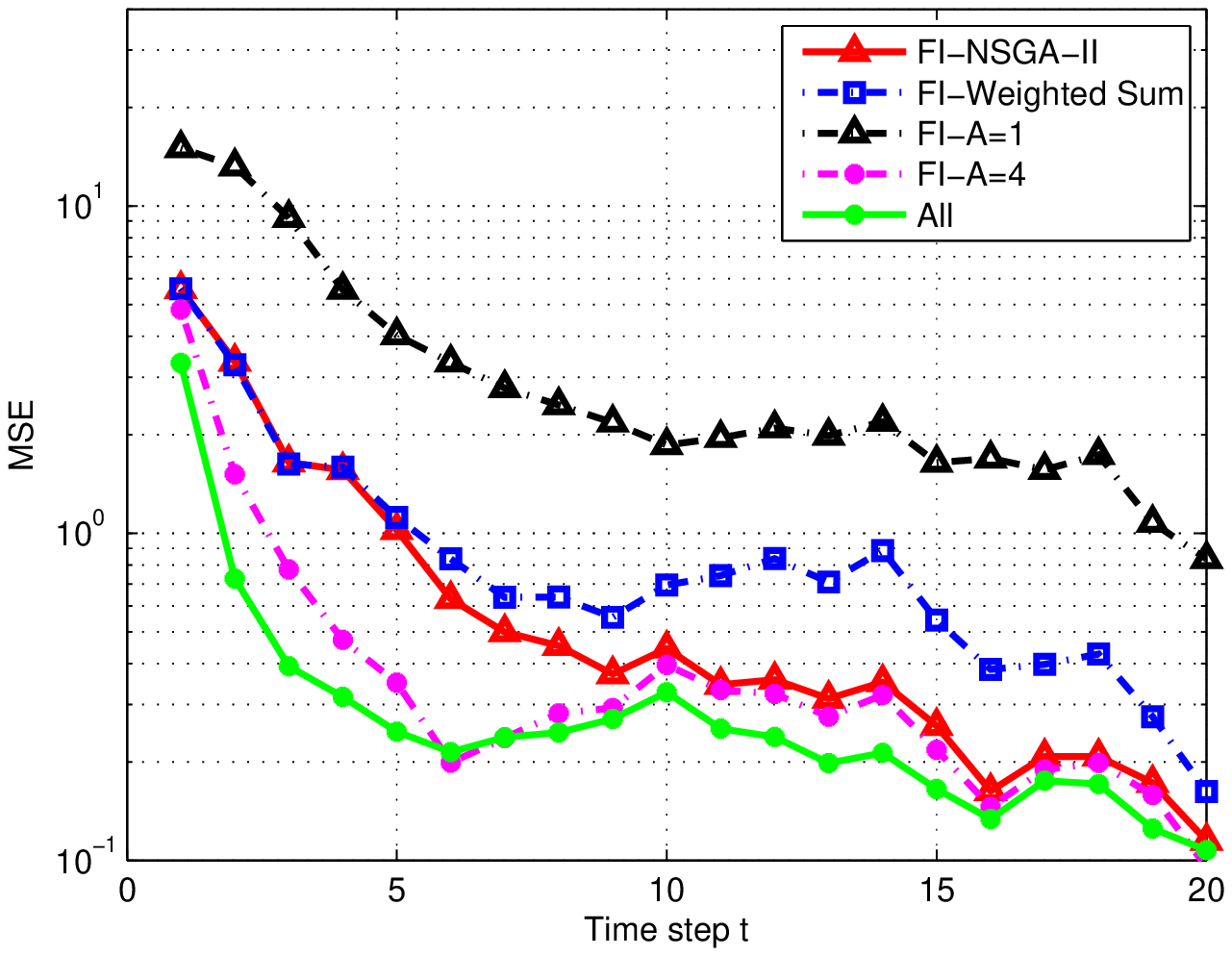}
\label{subfig:MOP_WS_FI}}
\caption{Tracking performance for MOP with NSGA-II, convex relaxation, and weighted sum methods (a) MSE for MIUB; (b) MSE for FI.}
\label{fig:MOP_WS}
\end{center}
\end{figure*}

In this section, we conduct some simulation experiments to investigate the performance of the multiobjective optimization method. The WSN considered in this subsection is the same as shown in Fig. \ref{fig:wsn} in Section \ref{Subsec:criteria_sims}, and the system parameters are also the same as Section \ref{Subsec:criteria_sims}. Note that, for NSGA-II, the population size is chosen as $P = 100$. 
We choose the number of generations according to the diversity metric introduced in \cite{NSGA-II}. The diversity metric measures the extent of spread achieved among the obtained solutions, which is defined as
\begin{equation}
\Delta = \dfrac{d_f^{(E)}+d_l^{(E)}+\sum_{i=1}^{N-1} |d_i^{(E)}-\bar{d}^{(E)}|}{d_f^{(E)}+d_l^{(E)}+(N-1)\bar{d}^{(E)}}
\end{equation}
where $d_f^{(E)}$ and $d_l^{(E)}$ are the Euclidean distances between the extreme solutions and the boundary solutions of the obtained nondominated set. 
We observe that for both FISS and MIUBSS, the diversity metric converges after 100 generations for all the 20 time steps. Thus, in our simulation experiments, we set the number of generations as $G=100$. 
Also, before running NSGA-II, we include the two extreme solutions, i.e, all zero and all one solutions to the initial population. 



\paragraph{Pareto optimal front} In Fig. \ref{fig:front}, we present the Pareto optimal front for our MOP obtained using NSGA-II, where Fig. \ref{subfig:front_FI} is for FISS and Fig. \ref{subfig:front_MIUB} shows the result for MIUBSS. It is interesting to note that at the end of $G$ generations, NSGA-II yields $N+1$ different solutions on the Pareto-optimal front where each solution corresponds to the optimal selection of $A$ sensors out of $N$ sensors where $A \in \{0,1,\ldots,N\}$. We know from \eqref{eq:JSA}, \eqref{eq:JSQ} and Table \ref{tab:sensor_index} that the distance between the target and the sensor plays a more important role than the sensing probability for FISS. At time step $t=3$, the target is relatively close to sensor 8, and sensor 8 itself is able to achieve significant FI gain. At time step $t=6$ the target is not relatively close to any of the sensors in the network and the fusion center has relatively large uncertainty about the target location. Thus, the Pareto front for FISS at $t=3$ is steeper than that at $t=6$. 
However, compared with FISS, MIUBSS prefers the sensors with high sensing probability and selects more sensors, so that the Pareto front of MIUBSS at $t=3$ or $t=6$ is not as steep as that for FISS. Moreover, we observe that the compromise solution and the knee point solution are identical when the Pareto front is relatively steep. 

\paragraph{Solution selection method} The solution, i.e., the sensor selection strategy, that we choose from the Pareto optimal front determines the overall tracking performance. In Fig. \ref{fig:MOP_SolS}, we compare the average number of active sensors\footnote{We show the number of active sensors (the selected sensors) to investigate the energy cost of each solution selection method, because selecting more sensors for data transmission incurs more energy cost.} at each time step of tracking and the MSE performance using the knee point solution \eqref{eq:kp_sol} and the compromise solution \eqref{eq:comp_sol} with MIUBSS and FISS under the MOP framework. We observe similar results for MIUB and FISS that the knee point solution always selects one sensor for target tracking, and thus gives poorer MSE performance. 
However, the sensor selection strategy using the compromise solution in \eqref{eq:comp_sol} selects the sensors which balance the tradeoff between the performance gain (MIUB and FI) and the total number of selected sensors. Thus, in the rest of our simulations, we use the compromise solution to choose the sensor selection strategy from the Pareto optimal front.

Recall the results shown in Fig. \ref{fig:onesensor} and Fig. \ref{fig:MI_MIUB} that MIUBSS selects more reliable sensors when the number of sensors to be selected is given. Furthermore, Fig. \ref{fig:MOP_SolS} shows that when the number of sensors to be selected is not known, MIUBSS tends to select more sensors than FISS under the MOP framework, such that the MSE performance of MIUBSS is better than FISS.

\paragraph{NSGA-II, convex optimization, and weighted sum methods} In Fig. \ref{fig:MOP_WS}, we compare the tracking performance based on NSGA-II and (\ref{eq:comp_sol}) with the convex relaxation based sensor selection method similar to \cite{Joshi:tsp09,engin_bit_allocation} which always chooses $A$ sensors out of $N$ sensors at each time step of tracking. 
We apply the convex relaxation method to select the minimum and maximum number of sensors selected by NSGA-II with compromise solution in Fig. \ref{fig:MOP_SolS}. With the minimum number of sensors, the convex relaxation based sensor selection method gives poor tracking performance. On the other hand, selecting the maximum number of sensors or all the sensors through convex relaxation method negligibly improves the MSE performance compared to the MOP approach. 
Thus, compared to the convex relaxation method, the multiobjective optimization method gives satisfactory tracking performance while saving in terms of the number of selected sensors with both MIUBSS and FISS. We also compare the MSE performance of the MOP framework with the weighted sum approach where the sensor selection scheme chooses those sensors which minimize the summation of both objectives, i.e. $w_1 f_1(\boldsymbol{\alpha}_t) + (1-w_1) f_2(\boldsymbol{\alpha}_t)$ with $w_1 = 0.5$.  
Simulation results show that for MIUBSS (Fig. \ref{subfig:MOP_WS_MIUB}), the NSGA-II method obtains similar MSE performance with weighted sum method, while for FISS (Fig. \ref{subfig:MOP_WS_FI}), the weighted sum method achieves much worse MSE performance.


\begin{figure}[tb]
\centerline{ \begin{tabular}{c}
\includegraphics[width=.7\columnwidth,height=!]{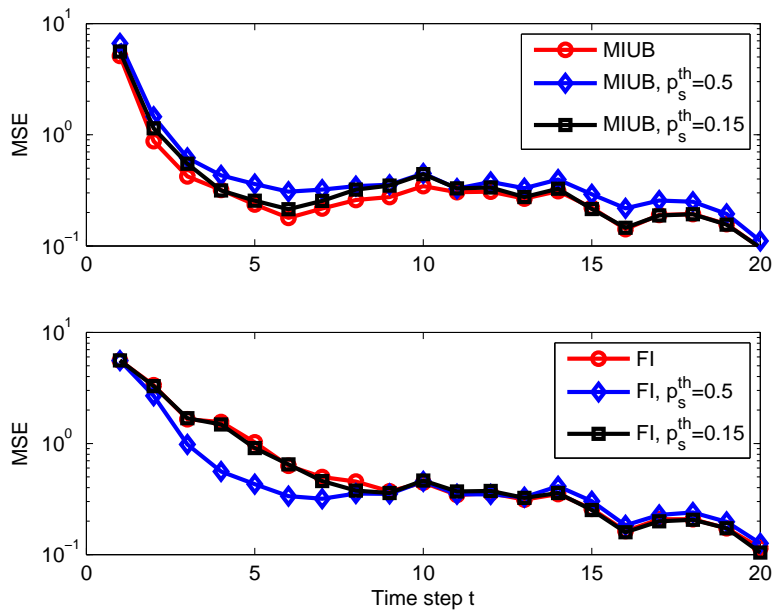} \end{tabular} }
      \caption{Turn off sensors with relatively low sensing probabilities.}
  \label{fig:MOP_Trunoff}
\end{figure}

\begin{figure}[tb]
\centerline{ \begin{tabular}{c}
\includegraphics[width=.7\columnwidth,height=!]{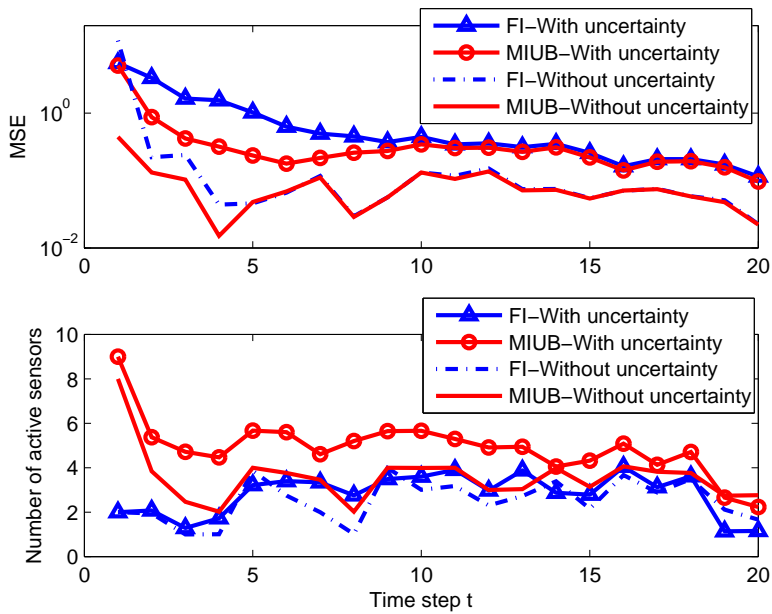} \end{tabular} }
      \caption{Comparison the performance for WSNs with and without uncertainty.}
  \label{fig:Without_Uncertainty}
\end{figure}

\begin{figure}[tb]
\centerline{ \begin{tabular}{c}
\includegraphics[width=.7\columnwidth,height=!]{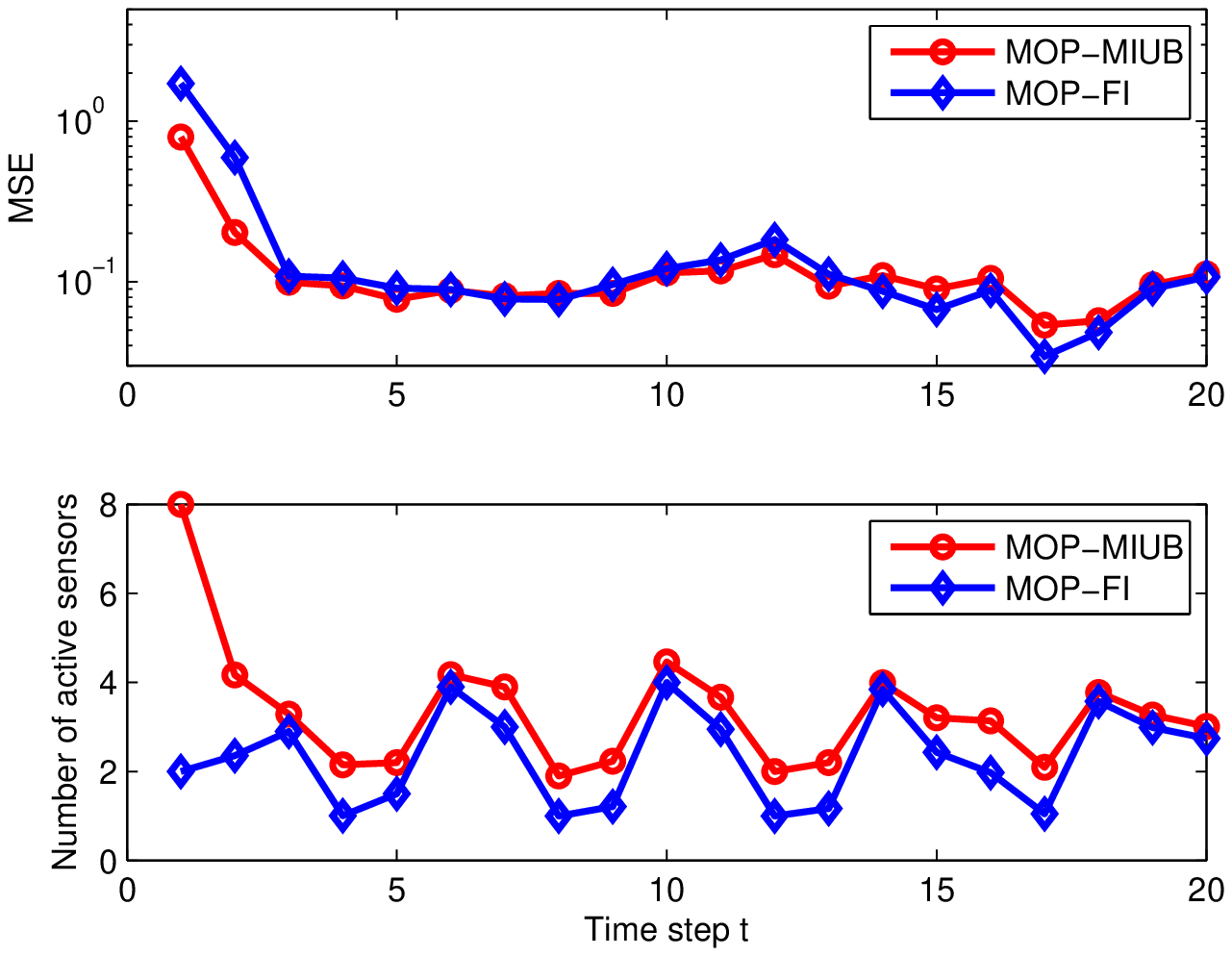} \end{tabular} }
      \caption{Tracking performance of MIUB and FI, sensors' sensing probabilities are reversely deployed.}
  \label{fig:MOP_MIUB_FIM_reverse}
\end{figure}




\paragraph{A Naive strategy} We consider a naive sensor selection method in which the fusion center turns off the sensors with relatively low sensing probabilities before sensor selection. In Fig. \ref{fig:MOP_Trunoff}, we present the results when the fusion center turns off the sensors whose sensing probabilities are lower than some threshold $p_s^{th}$, where $p_s^{th} = 0.5$ and $p_s^{th} = 0.15$ are considered. Note that, for the WSN in Fig. \ref{fig:wsn}, the sensors that are relatively close to the target will be turned off because they have low sensing probabilities. As shown in the previous results, MIUBSS prefers to select more reliable sensors, and FISS selects the sensors that are close to the target. Turning off sensors before selection performs worse for MIUBSS because it reduces the selection alternatives, and it performs better for FISS because more reliable sensors are selected when the closest sensors with low sensing probabilities are no longer available. 

\paragraph{Comparison with the performance when there is no uncertainty} In Fig. \ref{fig:Without_Uncertainty}, we present the target tracking performance when the sensors are all reliable, e.g., $p_s=1$ for all the sensors, and compare with the results with uncertain observations. 
We observe that, with uncertain observations, both FISS and MIUBSS achieve worse MSE performance though they both tend to select more sensors. 
Moreover, compared with FISS, MIUBSS selects many more sensors with uncertain observations, and therefore achieves better MSE performance.


\paragraph{WSN with another instance of sensing probabilities} In Fig. \ref{fig:MOP_MIUB_FIM_reverse}, the sensors' sensing probabilities are distributed in a reverse manner as compared with Fig. \ref{fig:wsn}, i.e., the sensors that are around the target track have relatively high sensing probabilities. In this condition, MIUBSS and FISS select similar number of sensors with similar MSE performance. The reason is that, under this scenario, MIUBSS and FISS both select the sensors around the target track with high sensing probabilities. 
We also have conducted experiments for the following scenarios: 1) the sensors' sensing probabilities are all uniformly distributed between 0 and 1; 2) the sensor measurements have higher noise; and 3) the sensor measurements are quantized to 3 bits, since the results do not provide any new insights, we do not show the results in the paper.

\section{Conclusion}
\label{Sec:Concl}
In this paper, we have proposed a multiobjective optimization method for the sensor selection problem in an uncertain wireless sensor network (WSN) for target tracking. 
We have considered the three performance metrics, Fisher information (FI), mutual information (MI), and mutual information upper bound (MIUB), as objective functions in characterizing the estimation performance for the multiobjective optimization problem (MOP). Numerical results show that the MIUB based selection scheme (MIUBSS) selects more reliable sensors compared with the FI based selection scheme (FISS) while saving computational cost compared with the MI based selection scheme (MISS). Furthermore, for the MOP framework, we have shown that the compromise solution on the Pareto front of the MOP achieves good estimation performance while obtaining savings in terms of the number of selected sensors.

In this work, we were interested in finding the sensor selection strategy with a multiobjective optimization method in uncertain WSNs. Future work will consider the application of the multiobjective optimization method in the multitarget tracking problem in uncertain WSNs. 

\bibliography{journal}
\bibliographystyle{IEEEtran}

\end{document}